\documentclass[12pt]{article}
\begin{document}

\title{\bf Constrained analytical interrelations in neutrino mixing}

\author{{\bf Biswajoy Brahmachari}\footnote{biswa.brahmac@gmail.com} \\ 
Department of Physics, Vidyasagar Evening College\\
39, Sankar Ghosh Lane, Kolkata 700006\\
and\\
{\bf Probir Roy}\footnote{probirrana@gmail.com} \\
Saha Institute of Nuclear Physics\\
Block AF, Bidhannagar, Kolkata 700064\\
and\\
CAPPS, Bose Institute, Kolkata 700091
}

\date{November 2015}

\maketitle

\begin{center}
\thanks{To be published in Springer Proceedings in the Physics Series\\ under the heading  of the XXI DAE-BRNS Symposium (Guwahati, India) }
\end{center}

\abstract{
 Hermitian squared mass matrices of charged leptons and 
light neutrinos in the flavor basis are studied under general additive 
lowest order perturbations away from the tribimaximal (TBM) limit in 
which a weak basis with mass diagonal  charged leptons is chosen. Simple 
analytical expressions are found for the three measurable TBM-deviants 
in terms of perturbation parameters appearing in the neutrino and charged 
lepton eigenstates in the flavor basis. Taking unnatural cancellations 
to be absent and charged lepton perturbation parameters to be 
small, interrelations are derived among masses, mixing angles and the amount 
of CP-violation.
}

\section{Introduction}
Recent 
global fits~\cite{ref3n,ref1,ref3,ref2,ref1-2} 
of the pairwise mixing angles \cite{ref11}, appearing in $U_{PMNS}$, yield 
the $3\sigma$ ranges
$31^0 < \theta_{12} < 36^0$, $36^0 < \theta_{23} < 55^0$ 
and $7.2^0 < \theta_{13} < 10^0$. On the other hand it is 
established \cite{ref3,ref2} that
$7.00< \Delta_{21}(10^5~eV^{2}) < 8.09$ and  $2.195 
< [\Delta_{32}(10^3~eV^{2})>0]< 2.625$ 
or $-2.649 < [\Delta_{32}(10^3~eV^{2})<0]<-2.242$, 
where $\Delta_{ij}=|m_{\nu i}|^2-|m_{\nu j}|^2$ and $m_i,~(i=1,2,3)$ are 
neutrino mass eigenvalues. 
Moreover, cosmological
observations, strengthened by recent data from the PLANCK satellite, claim 
\cite{ref4} that $\sum_{i} |m_{\nu i}| < 0.23 ~eV$.
Let $U_t$ be the unitary transformation 
that diagonalizes the hermitian squared mass matrix $M^\dagger_t M_t$ in
the flavor basis of a fermion of type $t$. Then ${U_u, U_d, U_\ell}$ can all be taken 
to show a hierarchical pattern, whereas $U_\nu$ seems governed by a different principle. 

\section{Perturbation theory of neutrino states} 
Let us henceforth use the superscript zero everywhere to denote
the TBM limit. Suppose, in this limit, we choose a basis with mass 
diagonal charged leptons. Let us also encapsulate neutrino masses and mixing angles 
in the complex symmetric Majorana mass matrix 
$M_{\nu f} = U_\nu^* M_\nu U_\nu^\dagger$ in the flavor 
basis. The 
TBM pattern obtains with $\theta^0_{12}=\sin^{-1}\sqrt{1/3} \sim 35.3^\circ$, 
$\theta^0_{23}=\sin^{-1} \sqrt{1/2}=45^\circ$ and $\theta^0_{13}=0$.
The measurement of a significantly nonzero 
value of $\theta_{13}$ has been a major experimental advance 
recently~\cite{ref92,ref93,ref91,ref9}.

 We add small general
perturbations to the TBM limits of hermitian squared mass 
matrices ${M_\ell}^\dagger M_\ell$ and 
$M^\dagger_{\nu f} M_{\nu f}$. We 
characterize them~\cite{us} by a set of small parameters $\{ \epsilon^{\nu,\ell} \}$. All members 
of the subset 
$\{\epsilon^\nu\}$ in the neutrino sector are taken to be typically of magnitude 
$\sim s_{13} \equiv \sin \theta_{13} \sim 0.16$, i.e. of the order of 16$\%$ (or thereabouts) 
of the 
unperturbed quantities. 

In the TBM limit ~\cite{ref5,ref51,ref7,ref6,ref8,ref81,ref52,ref4}, 
one has
\begin{equation}
{U^0_\nu}^\dagger ~{M^0_{\nu f}}^\dagger M^0_{\nu f} ~U^0_\nu={\rm diag}~
(|m^0_{\nu 1}|^2,|m^0_{\nu 2}|^2,|m^0_{\nu 3}|^2),
\end{equation}
where
\begin{equation}
U^0_\nu= \pmatrix{\sqrt{2/3} & \sqrt{1/3} & 0 \cr
       -\sqrt{1/6} & \sqrt{1/3} & \sqrt{1/2} \cr
       \sqrt{1/6} & -\sqrt{1/3} & \sqrt{1/2}}.
\end{equation}
The normalized eigenvectors of ${M^0_{\nu f}}^\dagger 
M^0_{\nu f}$ are the columns of $U^0_\nu$\cite{ref102,ref12,ref121,ref10,spar,ref101}. With 
the perturbation added, $M_\ell=M^0_\ell+M^\prime_\ell$ and
\begin{equation}
M^\dagger_\ell M_\ell=U^\dagger_\ell M^\dagger_{\ell f} M_{\ell f} U_\ell.
\end{equation}
Here $M^0_\ell=M^0_{\ell f}$ and $M^\prime_\ell=U^\dagger_\ell M^\prime_{\ell f} U$.
We similarly decompose $M_{\nu f}$ into two parts; $M_{\nu f}=M^0_{\nu f}+M^\prime_{\nu f}$, where 
$M^0_{\nu f}$ obeys the 
TBM conditions while
$M^\prime_{\nu f}$ violates them. Retaining  only terms linear in
the elements of  
$M^\prime_{\ell f}$ and $M^\prime_{\nu f}$, the 
ith eigenvectors of $M^\dagger_{\nu f}M_{\nu f}$ 
on one hand and of $M^\dagger_{\ell f}M_{\ell f}$ on the
other can be written in a compact 
notation with the perturbation parameters $\epsilon^{\nu,\ell}_{ik}$ (for $i,k=1,2,3$) as
\begin{equation}
|\psi^{\nu,\ell}_{i}\rangle_{f} = |\psi^{0 \nu,\ell}_{i}\rangle_{f} 
+\sum_{k \ne i}\epsilon^{\nu,\ell}_{ik} | \psi^{0 \nu,\ell}_{k} \rangle_{f} 
+ O(\epsilon^2).\label{7a}
\end{equation}
\begin{eqnarray}
\epsilon^{\nu,\ell}_{ik} &=& -{\epsilon^{\nu,\ell}}^*_{ki}=
(|m^0_{\nu,\ell i}|^2-|m^0_{\nu,\ell k}|^2)^{-1}p^{\nu,\ell}_{ki},\label{7b}\\
p^{\nu,\ell}_{ik} &=& \langle \psi^{0 \nu,\ell}_i|{M^0_{\nu,\ell}}^\dagger M^\prime_{\nu,\ell} 
+ {M^\prime_{\nu,\ell}}^\dagger M^0_{\nu,\ell}|\psi^{0 \nu,\ell}_k \rangle. \label{7c}
\end{eqnarray} 
Note that (\ref{7b}) and (\ref{7c}) have been written in the
mass basis utilizing the fact that $\epsilon^{\nu,\ell}_{ik}$ 
and $p^{\nu,l}_{ik}$ are the same
in either basis. On the other hand, the LHS of (\ref{7a}) for $i=1,2,3$ 
can be identified 
with the three
corresponding columns of $U_{\nu,\ell}$, i.e. 
\begin{equation}
U_{\nu,\ell} = (|\psi^{\nu,\ell}_1\rangle_{f}~|\psi^{\nu,\ell}_2\rangle_{f} 
~|\psi^{\nu,\ell}_3\rangle_{f}). \label{8}
\end{equation}

 Let us define the Majorana phase matrix\cite{mohapatra} $K={\rm diag}.
(1,e^{i \alpha_{21}/2},e^{i \alpha_{32}/2})$.
Then 
\begin{eqnarray}
&& U_{PMNS}=U^\dagger_\ell U_\nu K. \label{A} 
\end{eqnarray}
The RHS of (\ref{A}) can be identified with the form of $U_{PMNS}$ in the 
PDG convention \cite{ref11}. The Majorana phase matrix cancels out in the above identification.
Moreover, one is led to four independent constraints\cite{us}:
\begin{eqnarray}
{\rm Im}~\epsilon^\nu_{12} &=& O(\epsilon^2), \label{B} \\
{\rm Im}~(\epsilon^\nu_{13}-\sqrt{2} \epsilon^\nu_{23}) &=& O(\epsilon^2), \label{C}\\
{\rm Im}~\epsilon^l_{23} &=& O(\epsilon^2), \label{D} \\
{\rm Im}~(\epsilon^l_{12}-\epsilon^l_{13}) &=& O(\epsilon^2). \label{E}
\end{eqnarray}
In addition, the following expressions\footnote{Note that $\sqrt{2}~c_{12}+s_{12}
=\sqrt{3}+O(\epsilon^2)$
and $c_{23}+s_{23}=\sqrt{2} + O(\epsilon^2)$ are automatic}, 
which are linear 
in the $\epsilon$ parameters, emerge for the
three measurable deviants from tribimaximal mixing:
\begin{eqnarray}
&& c_{12}-\sqrt{2 \over 3}=\sqrt{1 \over 2}\left(\sqrt{1 \over 3} -s_{12} \right)
=\sqrt{1 \over 3} ~\epsilon^\nu_{12} 
- \sqrt{1 \over 6}\left( \epsilon^l_{12}-\epsilon^l_{13} \right), \label{F} \\
&& c_{23}-s_{23}=-\sqrt{2 \over 3}\left(\epsilon^\nu_{13}-\sqrt{2} 
~\epsilon^\nu_{23}\right)-\sqrt{2}~ \epsilon^l_{23}, \label{G} \\
&& s_{13}~e^{i \delta_{CP}}=-\sqrt{1 \over 3}~\left(\sqrt{2}~ \epsilon^\nu_{13} 
+ \epsilon^\nu_{23}\right)
+\sqrt{1 \over 2}\left(\epsilon^l_{12}+\epsilon^l_{13}\right). \label{H}
\end{eqnarray}

 Let us now take the perturbing mass matrices for neutrinos and charged leptons, 
with respective complex mass dimensional parameters $\mu_{ij}=\mu_{ji}$ and $\lambda_{ij}$, as
\begin{equation}
M^\prime_{\nu f}=
\pmatrix{ \mu_{11} & \mu_{12} & \mu_{13} \cr
          \mu_{12} & \mu_{22} & \mu_{23} \cr
          \mu_{13} & \mu_{23} & \mu_{33}},~~~~
M^\prime_{\ell f}=\pmatrix{\lambda_{11} & \lambda_{12} & \lambda_{13} \cr
                                  \lambda_{21} & \lambda_{22} & \lambda_{23} \cr
                                  \lambda_{31} & \lambda_{32} & \lambda_{33}}
=M^\prime_\ell+O(\epsilon^2) \label{K}   
\end{equation}
Eqn. (\ref{7b}) and Eqn. (\ref{7c}) 
lead to
\begin{eqnarray}
\epsilon^l_{12} &=& ({m^0_e}^2-{m^0_\mu}^2)^{-1}~(m^0_\mu \lambda_{21} + m^0_e \lambda^*_{12}), \label{M}\\
\epsilon^l_{23} &=& ({m^0_\mu}^2-{m^0_\tau}^2)^{-1}~(m^0_\tau \lambda_{32} + m^0_\mu \lambda^*_{23}), \label{N}\\
\epsilon^l_{13} &=& ({m^0_e}^2-{m^0_\tau}^2)^{-1}~(m^0_\tau ~\lambda_{31}+m^0_e{\lambda^*_{13}}). \label{O}
\end{eqnarray}
Because of the hierarchical nature of charged lepton masses,
 (\ref{D}) and (\ref{E}) can only be satisfied, without unnatural cancellations, 
by $\lambda_{12},\lambda_{21},\lambda_{13},\lambda_{31},\lambda_{32}$ and $\lambda_{23}$ all
being real to order $\epsilon$. That immediately implies
\begin{equation}
{\rm Im} ~\epsilon^l_{12}=O(\epsilon^2)={\rm Im}~ \epsilon^l_{13},\label{P}
\end{equation}
\begin{equation}
\tan \delta_{CP}
={3 ~{\rm Im}~\epsilon^\nu_{23} + O(\epsilon^2)
\over
{\rm Re}~[\sqrt{2} ~\epsilon^\nu_{13} + \epsilon^\nu_{23}]
-\sqrt{3 / 2} ~ {\rm Re}~[\epsilon^l_{12} + \epsilon^l_{13}] + O(\epsilon^2)}, 
\label{Q}
\end{equation}
\begin{equation}
J=-{1 \over \sqrt{6}}~{\rm Im}~\epsilon^\nu_{23} +O(\epsilon^2).\label{Q1}
\end{equation}
 For neutrinos, the relevant off-diagonal elements of
$M^\prime_\nu$ are
\begin{eqnarray}
(M^\prime_\nu)_{12}
&=& {1 \over 3 \sqrt{2}}(2 \mu_{11}+\mu_{12}-\mu_{13}-\mu_{22}+2 \mu_{23}-\mu_{33}), \label{R}\\
(M^\prime_\nu)_{23}
&=& {1 \over \sqrt{6}}(\mu_{12}+ \mu_{13}+ \mu_{22} - \mu_{33}), \label{T}\\
(M^\prime_\nu)_{13}
&=& {1 \over \sqrt{3}}(\mu_{12}+ \mu_{13}-{1 \over 2} \mu_{22} + {1 \over 2}
\mu_{33}). \label{S}
\end{eqnarray}
Each RHS above is nonzero if all the TBM conditions
are violated by $M^\prime_{\nu f}$\cite{us}. It is now convenient
to define
\begin{eqnarray}
&& \Delta^0_{ij} \equiv |m^0_{\nu i}|^2-|m^0_{\nu j}|^2,~~~~
a^\mp_{ij} \equiv m^0_{\nu i} \mp m^0_{\nu j}. \label{V}
\end{eqnarray}
Then (\ref{7b}), (\ref{7c}), (\ref{R}), (\ref{S}) and (\ref{T}) enable us to write
$\epsilon^\nu_{ik}$ in terms of the following combinations of elements of $M^\prime_{\nu f}$:
\begin{eqnarray}
 6\sqrt{2}~\Delta^0_{12}~
\pmatrix{{\rm i ~Im}~\epsilon^\nu_{12} \cr
{\rm Re}~\epsilon^\nu_{12}}
&&=
{a^\mp}^*_{21}~(2\mu_{11}+\mu_{12}-\mu_{13}-\mu_{22}+2 \mu_{23} - \mu_{33})
\mp c.c.,\label{W}\\
&& \nonumber\\
 2 \sqrt{6}~\Delta^0_{23}~
\pmatrix{{\rm i~Im}~\epsilon^\nu_{23} \cr
  {\rm Re}~ \epsilon^\nu_{23} }
&&= {a^\mp}^*_{32}(\mu_{12}+\mu_{13}+\mu_{22}-\mu_{33})
 \mp c.c., \label{Y} \\
&& \nonumber\\
 2 \sqrt{3}~ \Delta^0_{13}~
\pmatrix{ {\rm i~Im}~\epsilon^\nu_{13} \cr
 {\rm Re}~\epsilon^\nu_{13}}
&&= {a^\mp}^*_{31}(\mu_{12}+\mu_{13}-{1 \over 2}\mu_{22} 
+{1 \over 2}\mu_{33}) \mp c.c. \label{alpha}
\end{eqnarray}

\section{Implications and conclusion}
 
We can discuss \cite{us} the implications of (\ref{W}), (\ref{Y}), (\ref{alpha}) in
ongoing and forthcoming experiments\cite{ref15,ref13,ref161,ref131,ref17,
ref14,ref16}.
Once again, in the absence of unnatural cancellations,
(\ref{B}) and (\ref{W}) would require
$2 \mu_{11} + \mu_{12} -\mu_{13} - \mu_{22} + 2 \mu_{23} - \mu_{33}$
and $m^0_{\nu 2}$ to be real; the latter constrains the Majorana 
phase \cite{ref11} ${\alpha^0_{21}}$
to equal $0$ or $\pi$ in the TBM limit. These statements are valid neglecting $O(\epsilon^2)$
terms. Furthermore, (\ref{C}), (\ref{Y}) and (\ref{alpha}) would
require one of the following two conditions. {\bf Either,} one must
have {\bf condition 1:}
$m^0_{\nu 1}=m^0_{\nu 2}$, meaning $m^0_{\nu 1}=|m^0_{\nu 2}|$ plus $\alpha^0_{21}=0$,
and $\mu_{22}=\mu_{33}$ 
in which case,  
$\sqrt{2}~{\rm Re}~\epsilon^\nu_{23}={\rm Re}~\epsilon^\nu_{13}+O(\epsilon^2)$
and then, from  (\ref{G}), $c_{23}-s_{23}=-\sqrt{2}~\epsilon^l_{23}+O(\epsilon^2),
$i.e, $s_{23}=(1/\sqrt{2})(1+\epsilon^l_{23})+O(\epsilon^2)$; the latter
implies via (\ref{G}) and (\ref{N}) that any deviation from maximality in
the atmospheric neutrino mixing angle $\theta_{23}$ must come solely
from the 2-3 off-diagonal element in the charged lepton mass perturbation $M^\prime_l$
and is expected to be small since $\epsilon^l_{23}$  is 
scaled by $(m_\tau)^{-1}$, cf. (\ref{N}) .
{\bf Or}, what becomes necessary is {\bf condition 2:} 
$m^0_{\nu 3}-m^0_{\nu 1}$, $m^0_{\nu 3}-m^0_{\nu 2}$ as well as
$\mu_{12}+\mu_{13}$ and $\mu_{22}-\mu_{33}$ have to be real; this means that
the Majorana phase $\alpha^0_{21}$ and $\alpha^0_{32}$ in the TBM
limit are $0$ or $\pi$ and $\epsilon^\nu_{23}$ is
real, in which case, by virtue of (\ref{Q}) as well as (\ref{Q1}), 
$\sin ~\delta_{CP}=O(\epsilon^2)$ and $J=O(\epsilon^2)$ so that,
any observable CP-violation in neutrino oscillation experiments would vanish to the
lowest order of TBM violating perturbations. This is our statement on
two alternatives one of which is obligatory. Thus
$|s_{23}-{1 \over \sqrt{2}}|$ has to be significantly less than $s_{13}$ 
(though we are unable to make a precise prediction) or alternatively
$\sin \delta_{CP}$ as well as $J$ would be unobservably small.

%%%%%%%%%%%%%%%%%%%%%%%% Springer-Verlag %%%%%%%%%%%%%%%%%%%%%%%%%%
%
% BibTeX users please use
% \bibliographystyle{}
% \bibliography{}
%

%%%%%%%%%%%%% end %%%%%%%%%%%%%%%%%%%%%%%%%%%%%%%

 \end{document}